\documentclass[10pt,twocolumn]{article}
\usepackage[top=2cm, bottom=2cm, left=1.85cm, right=1.85cm]{geometry}
\usepackage{times}  %
\usepackage[sort&compress,numbers]{natbib}  %
\usepackage[hyphens]{url}
\usepackage{graphicx}  %
\usepackage[keeplastbox]{flushend}
\usepackage{graphicx} %
\usepackage{tikzsymbols}

\let\oldbibliography\thebibliography
\renewcommand{\thebibliography}[1]{%
\oldbibliography{#1}%
\setlength{\itemsep}{2pt}%
}
\usepackage[hang,flushmargin]{footmisc}

\usepackage[compact]{titlesec}
\titlespacing*{\section}{0pt}{*4}{4pt}
\titlespacing*{\subsection}{0pt}{*2}{2pt}
\usepackage{xspace}
\makeatletter
\def\url@leostyle{%
  \@ifundefined{selectfont}{\def\UrlFont{}}%
  {\def\UrlFont{}}%
}
\makeatother
\usepackage[hyphenbreaks]{breakurl}
\usepackage[bookmarks=true, bookmarksnumbered=true, colorlinks=true, linkcolor=linkcol, citecolor=citecol, urlcolor=urlcol, hypertexnames=true]{hyperref}
\definecolor{darkgreen}{RGB}{0, 100, 0}
\definecolor{linkcol}{rgb}{0.3,0,0}
\definecolor{citecol}{rgb}{0.3,0,0}
\definecolor{urlcol}{rgb}{0.3,0,0}

\usepackage[small,bf]{caption}
 \usepackage{multirow}

\usepackage{graphicx}
\usepackage{caption}
\usepackage{booktabs}                         
\usepackage{subfigure}
\usepackage[utf8]{inputenc}
\usepackage[english]{babel}
\usepackage{xspace}

\usepackage{enumitem}

\newcommand{\one}{({\em i}\/)\xspace}
\newcommand{\two}{({\em ii}\/)\xspace}
\newcommand{\three}{({\em iii}\/)\xspace}
\newcommand{\four}{({\em iv}\/)\xspace}

\def\eg{\emph{e.g.}\xspace}

\def\ie{\emph{i.e.}\xspace}

\def\vs{\emph{vs.}\xspace}

\newcommand{\pb}[1]{\vspace{0.75ex}\noindent{\bf \em #1}\hspace*{.3em}}
\newcommand{\paragraphbe}[1]{\vspace{0.75ex}\noindent{\bf \em #1}\hspace*{.3em}}

\newif
\ifcomment
\commenttrue

\ifcomment
\newcommand\gareth[1]{\textbf{\textcolor{red}{GT: #1}}	}
\newcommand\ar[1]{\textbf{\textcolor{blue}{AR: #1}}	}
\newcommand\ishaku[1]{\textbf{\textcolor{purple}{IA: #1}}	}
\newcommand\ignacio[1]{\textbf{\textcolor{green}{IC: #1}}	}
\newcommand\edc[1]{\textbf{\textcolor{brown}{EDC: #1}}	}

\else
\newcommand\gareth[1]{}
\newcommand\ar[1]{}
\newcommand\ishaku[1]{}
\newcommand\ignacio[1]{}
\newcommand\edc[1]{}
\fi

\colorlet{sharks}{blue}

\begin{document}

\sloppy

\graphicspath{{pleroma_plots/imc_plots}}

\title{\bf Exploring Content Moderation in the Decentralised Web:\\The Pleroma Case\thanks{Published in the Proceedings of the 17th International Conference on emerging Networking EXperiments and Technologies (ACM CoNext 2021).}}

\author{Anaobi Ishaku Hassan$^1$, Aravindh Raman$^2$, Ignacio Castro$^1$, Haris Bin Zia$^1$,\\
Emiliano De Cristofaro$^3$, Nishanth Sastry$^4$, and Gareth Tyson$^1$\\[0.5ex]
\normalsize $^1$Queen Mary University of London, $^2$Telefonica Research, $^3$University College London, $^4$University of Surrey
}

\date{}
\maketitle

\begin{abstract}

Decentralising the Web is a desirable but challenging goal.
One particular challenge is achieving decentralised content moderation in the face of various adversaries (\eg trolls).
To overcome this challenge, many Decentralised Web (DW) implementations rely on \emph{federation policies}.
Administrators use these policies to create rules that ban or modify content that matches specific rules.
This, however, can have unintended consequences for many users.
In this paper, we present the first study of federation policies on the DW, their in-the-wild usage, and their impact on users. 
We identify how these policies may negatively impact ``innocent'' users and outline possible solutions to avoid this problem in the future.

\end{abstract}

\section{Introduction}
\label{sec:intro}

The ``Decentralised Web'' (DW) is an evolving concept, which encompasses technologies  aimed at providing greater transparency, openness, and democracy on the web~\cite{guidi2018managing}.
Today, well-known social DW platforms include Pleroma, Mastodon (microblogging services), Hubzilla (cyberlocker), and PeerTube (video sharing platform).

Typically, individuals or organisations can install, own and manage their servers, also known as \textbf{instances}~\cite{Matteo, verge17}.
Instances are independent, and participants in the community must register with specific instances. For example, in the case of Pleroma (a microblogging service), a company could create an instance to provide a platform for employees to interact. To enable users across such instances to interoperate, \textbf{federation} protocols allow information and interactions to flow across DW instances to create a larger interconnected community.
This creates a physically distinct set of servers yet allows users to follow each other regardless of which instance they register with. 

A key selling point of the DW is the promotion of free speech outside of the remit of large tech companies.
Although appealing, this decentralised form of management  creates new challenges~\cite{content_warning}. 
For example, as popular centralised social platforms like Facebook and Twitter continue to clamp down on hateful and violent communities, some of these communities have migrated~\cite{dissenter} to DW instances where moderation and regulation are more difficult to enforce (\eg Gab \cite{gab,zannettou2018gab}).

In contrast to centralised services (\eg Twitter), 
DW moderation is usually performed on a per-instance basis.
Specifically, instance administrators enforce policies within their instance to moderate the content coming from other federated instances. 
For example, administrators of one instance can \textbf{reject} (\ie block) any material from other instances that match specific criteria.
This instance-based approach shifts the moderation responsibility to administrators who need to answer questions such as:
What policies should be applied, and to which instances?
How much effort should be put in moderation?
What is the collateral damage of the policies (\ie while a minority of users might cause a policy, this  will affect the rest of the ``innocent'' users of that same instance)?  

To explore these questions, we focus on one popular DW platform: \textbf{Pleroma}. 
In contrast to other DW microblogging platforms, Pleroma instances make their moderation policies public through an API.
We collect a large-scale dataset covering 5 months; this includes  1298 instances, 111k users, 24.5m posts, associated metadata, and, importantly, the 46 different policies imposed by the instances. 

We analyse the types of policies imposed by administrators. 
We find that that moderation affects the overwhelming majority of the users: 
97.7\% users and 97.8\% posts are impacted by policies. The \texttt{reject} action is most popular, affecting 86.2\% users and 88.5\% posts (see Section~\ref{section:pol}).
This brute-force policy blocks entire instances, even though only a subset of users might be misbehaving. 
To investigate the resulting collateral damage, we study users' toxicity using Google's Perspective API  (Section~\ref{section:coll}).
While toxic users are the likely target of instance blocking, 
we find that 95.8\% of the users blocked are not toxic. 
We conclude by proposing some strawman solutions to reduce collateral damage.

\section{Background}
\label{sec:pleroma}

\pb{Pleroma.}
Pleroma is a lightweight decentralised microblogging server implementation whose user-facing functionality is similar to Twitter. 
In contrast to a centralised social network, Pleroma is a federation of servers (aka \textbf{instances}), which interlink to share content.
Through these instances, users can register accounts and publish posts, which will appear on follower timelines. These followers can either be on the same instance or another (federated) instance.

\pb{Federation.}
We refer to users registered on the same instance as \textbf{local}, and users registered on different instances as \textbf{remote}. A user on one instance can follow another user on a separate instance.
Note that a user registered on their local instance does not need to register with the remote instance to follow the remote user. Instead, a user creates a single account with their local instance. 
When the user wants to follow a user on a remote instance, the local instance subscribes to the remote user on behalf of the local user, thereby federating with the remote instance. 
This process is implemented using an underlying subscription protocol (ActivityPub~\cite{activitypub}) that allows instances to federate with each other. 

\pb{Fediverse.} The resulting network of federated instances is referred to as the fediverse. The fediverse includes instances from Pleroma and instances from other platforms that Pleroma can federate with (\eg  Mastodon) because they support the same subscription protocol (\ie ActivityPub). Accordingly, Pleroma instances can federate and target its policies at non-Pleroma instances (\eg Gab from Mastodon). 

\pb{Policies.}
Instances in the fediverse federate with each other, and federated instances can target each other with policies.
Policies affect how instances federate with each other through 
different rule-action pairs. These allow certain actions to be executed when a post, user, or instance matches pre-specified criteria. We refer to each of these rule-action pairs within a policy as \textbf{actions} (\eg the \texttt{SimplePolicy} has multiple actions such as \texttt{media removal} and  \texttt{reject}). 

Note that some policies are in-built to the Pleroma software package. Instance administrators can enable (``switch on'') one or more policies.
Some of these policies are enabled by default when a new Pleroma instance bootstraps. 
Additionally, administrators can craft new policies if they require specific functionalities not covered by the in-built policies.

\section{Data Collection}
\label{section:data}

\paragraphbe{Instances.} 
Our measurement campaign covers the period between 16 December 2020 and 24 April 2021.
We first compile a list of Pleroma instances by crawling the directory of instances from ~\url{distsn.org} and~\url{the-federation.info}. 
We then capture the list of instances 
that each Pleroma instance has ever federated with
using the Peers API.\footnote{\url{<instance.uri>/api/v1/instance/peers}}

This includes both Pleroma and non-Pleroma instances (Pleroma can federate with any instance of the fediverse, see Section~\ref{sec:pleroma}).
In total, we identify 9969 instances, out of which 1534 are Pleroma and 8435 are non-Pleroma (\eg Mastodon).  

We then collect  metadata for each Pleroma instance every 4 hours  via their public APIs.\footnote{\url{<instance.uri>/api/v1/instance/}}
We obtain the number of users on the instance, the number of their followers, the number of posts, the version of Pleroma, whether the instance is accepting new registrations, the enabled policies, the applied policies as well as the instances targeted by these policies, and other meta information. 

From the 1534 Pleroma instances, we are able to gather data from 1298 instances (84.6\%). 
For the remaining 236 instances: 110 are not found (404 status code), 84 instances require authorisation for timeline viewing (403), 24 result in bad gateway (502), 11 in service unavailable (503), and 7 return gone (410).

\pb{User Timelines.}  Users have three timelines: 
\one~a \textit{home} timeline, with posts published by the accounts that the user follows (local and remote);
\two~a \textit{public} timeline, with all the posts generated within the local instance; 
and \three~the \textit{whole known network}, with \emph{all} posts that have been retrieved from remote instances that the local users follow. 
Note, the \textit{whole known network} is not limited to remote posts that a particular user follows; instead, it is the union of remote posts retrieved by all users on the instance. 
The whole known network timeline is an innovation-driven by the decentralised nature of instances: it allows users to observe and discover posts by remote users. 
Out of the 1,298 Pleroma instances, we gather all posts from 796 instances (119 instances had no posts, and the public timeline of the remaining 38.7\% instances was not reachable).
We gather data using the public Timeline API.\footnote{\url{<instance.uri>/api/v1/timelines/public?local=true}} 
This timeline covers all posts shared on each instance.
This allows us to collect 14.5M (including federated posts) public posts out of 24.5M posts, covering 91.7K users. 
Note that, from the 1,298 instances we are able to crawl, we discover a total of 111K unique users. 48.7\% of users published at least one post.

\pb{Harmful Classifications.} 
\label{class}
Generally, instance administrators target moderation policies/actions against other instances that have been perceived to post harmful content or violate their "Terms of Service." For any instance that has at least one \texttt{reject} action targeted against it (see Section~\ref{sec:policies:char}), we annotate all of its posts with harmful/non-harmful labels (15.8\% of all instances).
We annotate the posts using  Google's Perspective API~\cite{quick, automated}. 
The Perspective API scores text based on the perceived impact it might have on a conversation~\cite{perspective_works}. The scores represent the probability that a human annotator would reach the same conclusion and are between 0 to 1 for a range of attributes~\cite{perspective_model, jigsaw_perspective}.
In this paper, we classify posts across three attributes: toxicity, profanity, and sexually explicit content.
Perspective results have been found to be similar to human annotators~\cite{quick, compromise,dissenter, offensive},  it is widely used in production environments (\eg New York Times)~\cite{perspective_case} and it is continually maintained and updated~\cite{unintented}.

We label a \emph{post} as harmful if it receives a score of 0.8 or above in any of the three attributes (toxicity, profanity, and sexually explicit). 
This threshold is based on recommendations from the developers of the Perspective API~\cite{perspective_cards} and related literature~\cite{corel}.
Finally, we classify a \emph{user} as  harmful when the average of all the user's posts for any of the attributes (toxicity, profanity, sexually explicit) is greater or equal to 0.8. 

\section{Exploring Policies} 
\label{section:pol}

We begin by briefly characterising the types of policies used within Pleroma and the instances targeted by these policies. 

\begin{figure}[t]
	\includegraphics[width=\linewidth]{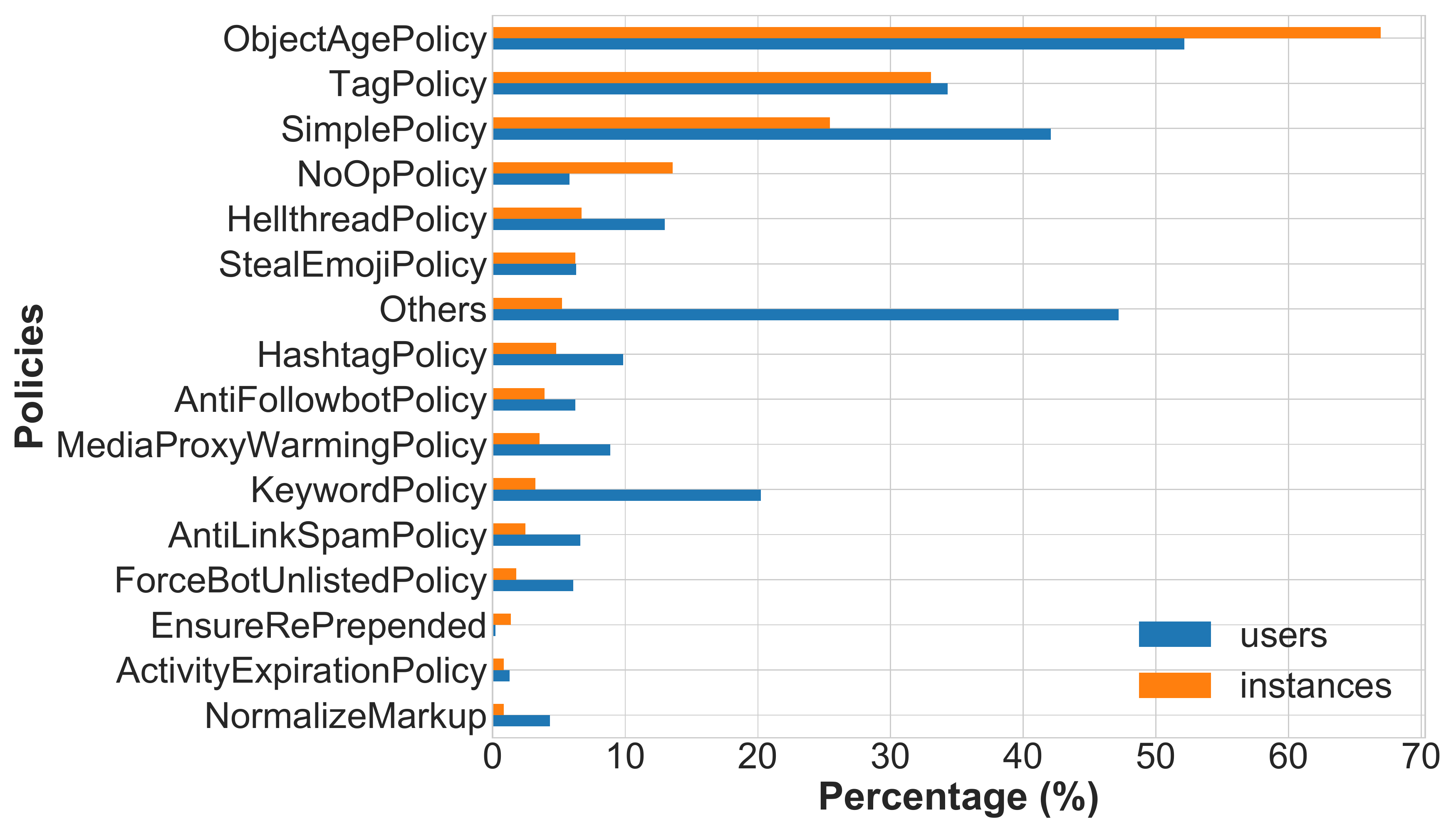}
	\caption{The top 15 policy types and the percentage of instances that use each policy (sorted by the percentage of instances). We also include the percentage of the global user population on the instances that use each policy. We represent the percentage of instances and users for all the less popular policies as ``Others.''}
	\label{fig:polices}
\end{figure}

\subsection{Characterising Policy Settings}
\label{sec:policies:char}

\pb{Overview of Policies.}
We are able to retrieve policy information from 91.9\% of Pleroma instances. The remaining 8.1\% of Pleroma instances do not expose their policy information.
These cover 46 unique policy types:
26 of these policies are included in the Pleroma software package, instance administrators have created the other 20. 
In general, administrators need to enable policies before they target them towards specific instances. 
However, we find the \texttt{ObjectAgePolicy and NoOpPolicy} enabled by default in the software package.

The policies we retrieve affect 97.7\% of the total users and 97.8\% of all posts. 
Figure~\ref{fig:polices} shows the distribution of the top 15  policy types enabled by the administrators across instances and the percentage of users signed up within those instances.

\pb{Popular Policies.}
Most common amongst the popular policies is the \texttt{ObjectAgePolicy} (on 66.9\% of instances). This policy allows admins to apply an action based on the age of a post regardless of the post's harmful/non-harmful nature. The default age threshold is 7 days but administrators are able to configure this as they choose. Possible actions under this policy includes~\one delist: removes the post from the public timelines; \two strip followers: removes followers from the recipient list; and \three reject: rejects the message entirely. As a default policy, this is enabled on any new installations of Pleroma starting from version 2.1.0.

The \texttt{TagPolicy}, applies policies to individual users based on tags but does not entirely stop the flow of any material between instances. For example, it allows marking posts from individual users as Not Safe For Work (NSFW). This policy is enabled on 33\% of instances. 
Finally, the \texttt{SimplePolicy} is enabled on 25.4\% of instances. 
The \texttt{SimplePolicy} is the most flexible policy, allowing admins to configure a range of actions on posts or instances that match certain criteria \eg the \texttt{reject} action blocks all connections from a given instance.

The remaining policies are less commonly encountered. 
For completeness, we briefly discuss a few of them here. The \texttt{HellthreadPolicy} is enabled on 6.7\% of the instances. This  de-lists/rejects a post when the number of user mentions exceeds a set threshold.
The \texttt{StealEmojiPolicy} whitelists instances to automatically download emojis from; this is enabled on 6.2\% of the instances.
Other less common policies we encounter include the \texttt{HashtagPolicy, AntiFollowBotPolicy, MediaProxyWarmingPolicy, KeywordPolicy} (see Appendix~\ref{append:pol} for a list of Pleroma in-built policies).

\begin{figure}[t]
  \includegraphics[width=\linewidth]{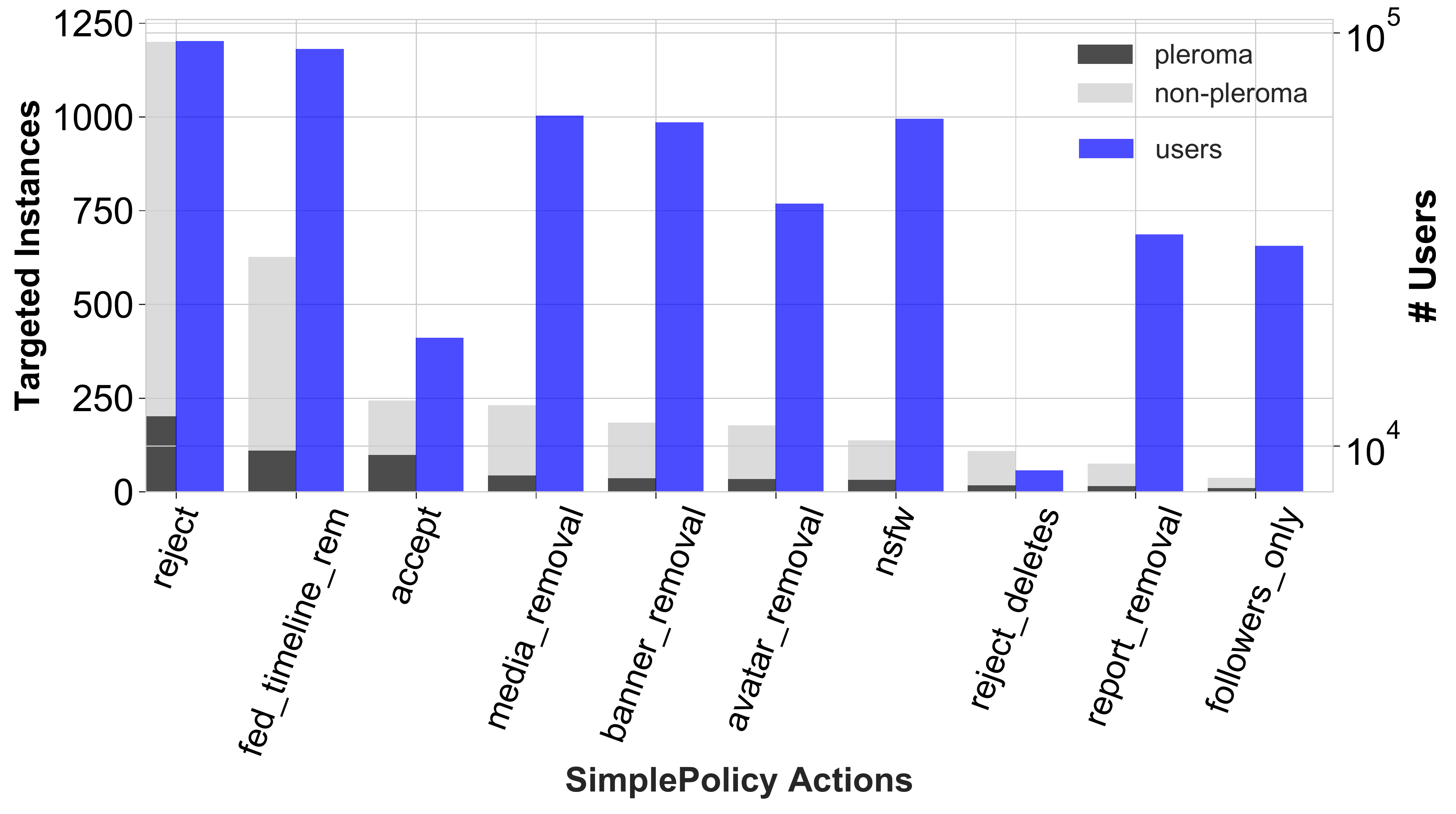}
  \caption{Number of instances targeted by \texttt{SimplePolicy} actions (Y-1). Instances are split into Pleroma and non-Pleroma instances. We also plot the number of users on the associated Pleroma instances (Y-2).}

  \label{fig:spolicy_breakdown}
\end{figure}

\begin{figure}[t]
 \centering
  \includegraphics[width=\linewidth]{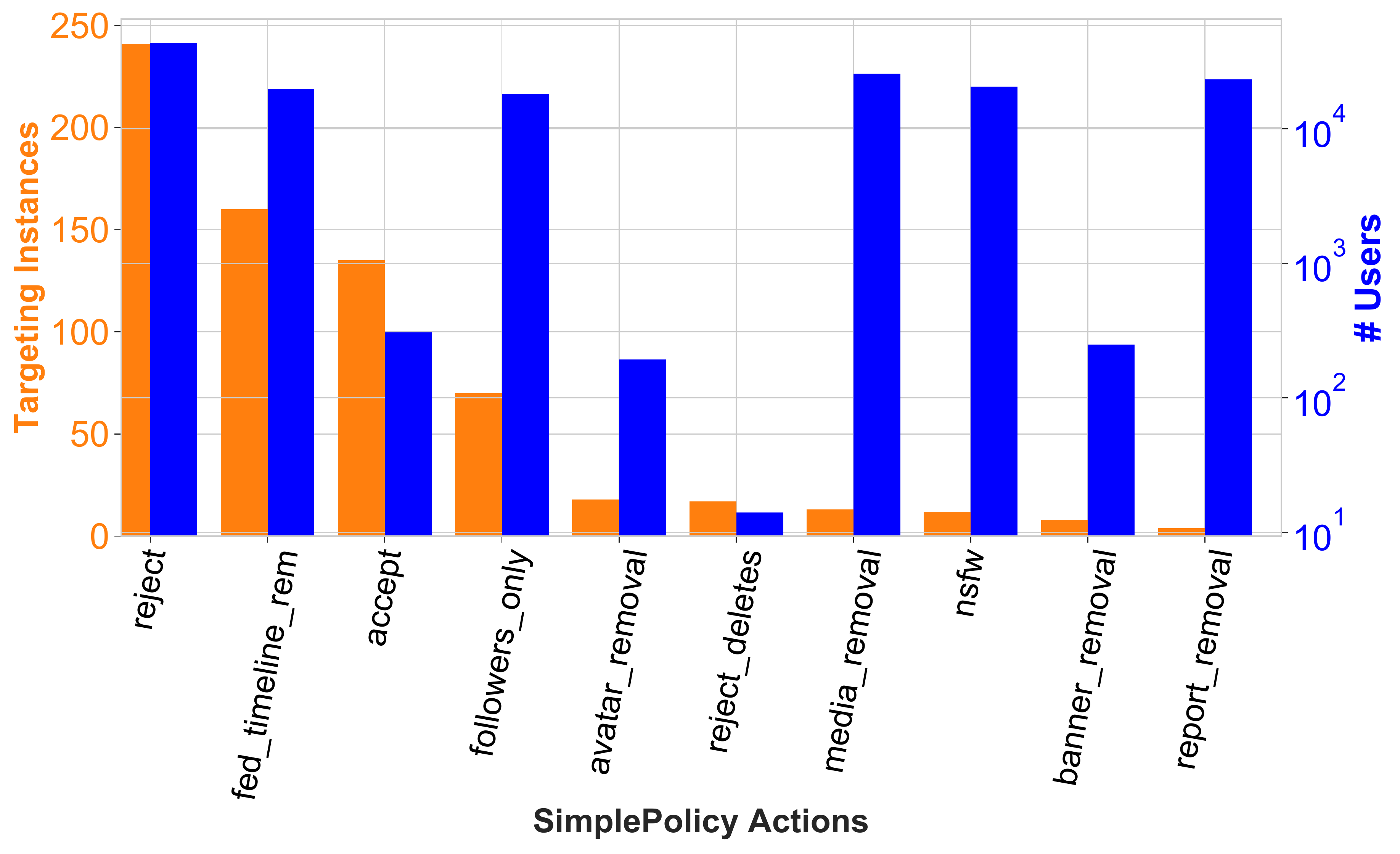}
  \caption{Bar plot showing the number of instances that have targeted other instances with the \texttt{SimplePolicy} actions. We also plot the number of users associated with the targeted instances.} 
  \label{fig:tageting_users}
\end{figure}

\pb{SimplyPolicy Breakdown.}
Due to the diversity of features available within the \texttt{SimplePolicy}, its reach  as well as its relevance in content moderation, we next inspect the most popular actions associated with the \texttt{SimplyPolicy}.
Figure~\ref{fig:spolicy_breakdown} presents a breakdown of the various actions used by instances with the \texttt{SimplePolicy} against Pleroma instances, as well as instances from other platforms of the fediverse (\eg Gab from Mastodon). The figure also shows the number of users signed up on these instances.
In contrast, Figure~\ref{fig:tageting_users} shows the number of Pleroma instances that have targeted other instances with the \texttt{SimplePolicy} actions and the number of users on these instances.

The figures reveal  a rich variety of policy actions. For example, the \texttt{media removal} action (which removes any media coming from targeted instances) 
is applied by 5.4\% of the instances, and this impacts 23.3\% of users.
The most popular and stringent is, however, the \texttt{reject} action.
Figure~\ref{fig:tageting_users} shows that the 73\% of instances that have the \texttt{SimplePolicy} enabled,  apply the reject action. 

We also notice 86.2\% of users and 88.7\% of posts are on instances that have been rejected by at least one other instance, and these rejected instances make up 80\% of all moderated instances. On a finer granularity, we see the \texttt{reject} action making up 62.8\% of all  moderation events while the sum of all the other (9) \texttt{SimplePolicy} actions make up the remaining 37.2\%. 
As the \texttt{reject} action is the most stringent, the most popular, and impacting a larger number of users, we focus our analysis on the \texttt{reject} action. Hence, we spend the next section exploring the instances that these \texttt{reject} actions are targeted against.

\subsection{Characterising Rejected Instances}

\begin{figure}[t]
  \includegraphics[width=0.5\textwidth]{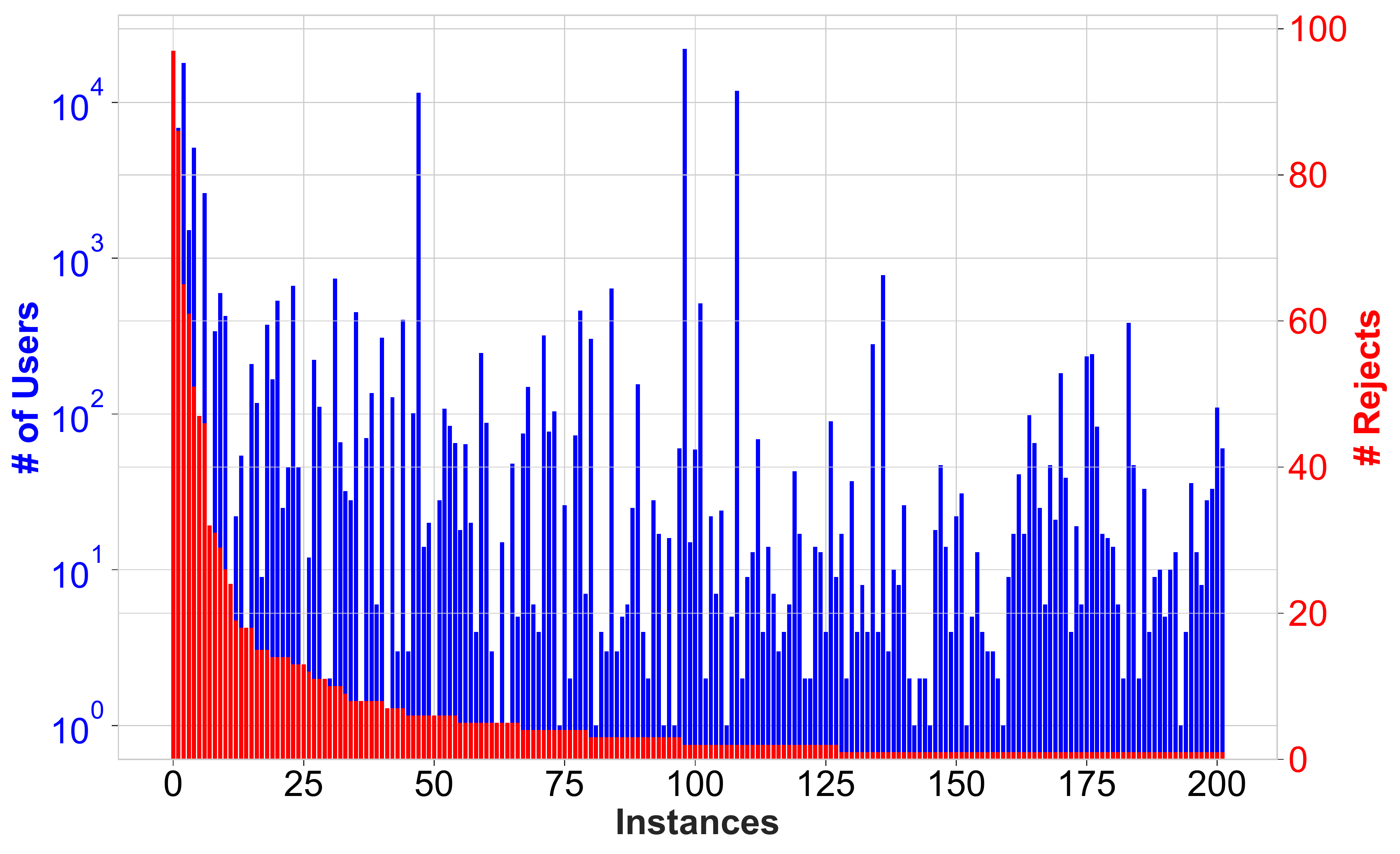}
  \caption{A bar plot showing all rejected Pleroma instances (X-axis) with their number of users and  the number of Pleroma instances that have rejected them (sorted by the number of rejects). }  
  \label{fig:rejected}
\end{figure}

\pb{Distribution of Reject Policies.}
Figure~\ref{fig:rejected} shows the number of \texttt{reject} actions targeting each Pleroma rejected instances. These instances represent only 15.5\% of Pleroma instances, however, they accumulate 86.2\% and 88.7\% of the total users and posts, respectively. 
Instances with more posts tend to receive a larger number of rejects: we find a weak correlation between the number of posts on an instance and the number of rejects (Spearman of 0.38).
Overall, we find 1,200 unique instances have been rejected at least once (202 Pleroma and 998 non-Pleroma).
The majority of these are targeted by only a small subset of instances though: 
86.8\% of these are rejected by fewer than 10 instances. However, we do see an ``elite'' set (5.4\%) of controversial Pleroma instances that gain in excess of 20 \texttt{reject} actions, led by \texttt{Freespeechextremist.com} (a proponent of free speech with 97 rejects). A variety of other types of instances also make up the top rejected list \eg \texttt{kiwifarms.cc} (well known for trolling, with 86 rejects), \texttt{spinster.xyz} (a feminist instance rejected 65 times) and \texttt{neckbeard.xyz} (blocked by another instance linking it to the LGBT community with 61 rejects). This ``elite'' set of rejected instances accounts for 33.6\% and 23.4\% of the total Pleroma users and posts, respectively.

\begin{figure}[t]
 \centering
  \includegraphics[width=\linewidth]{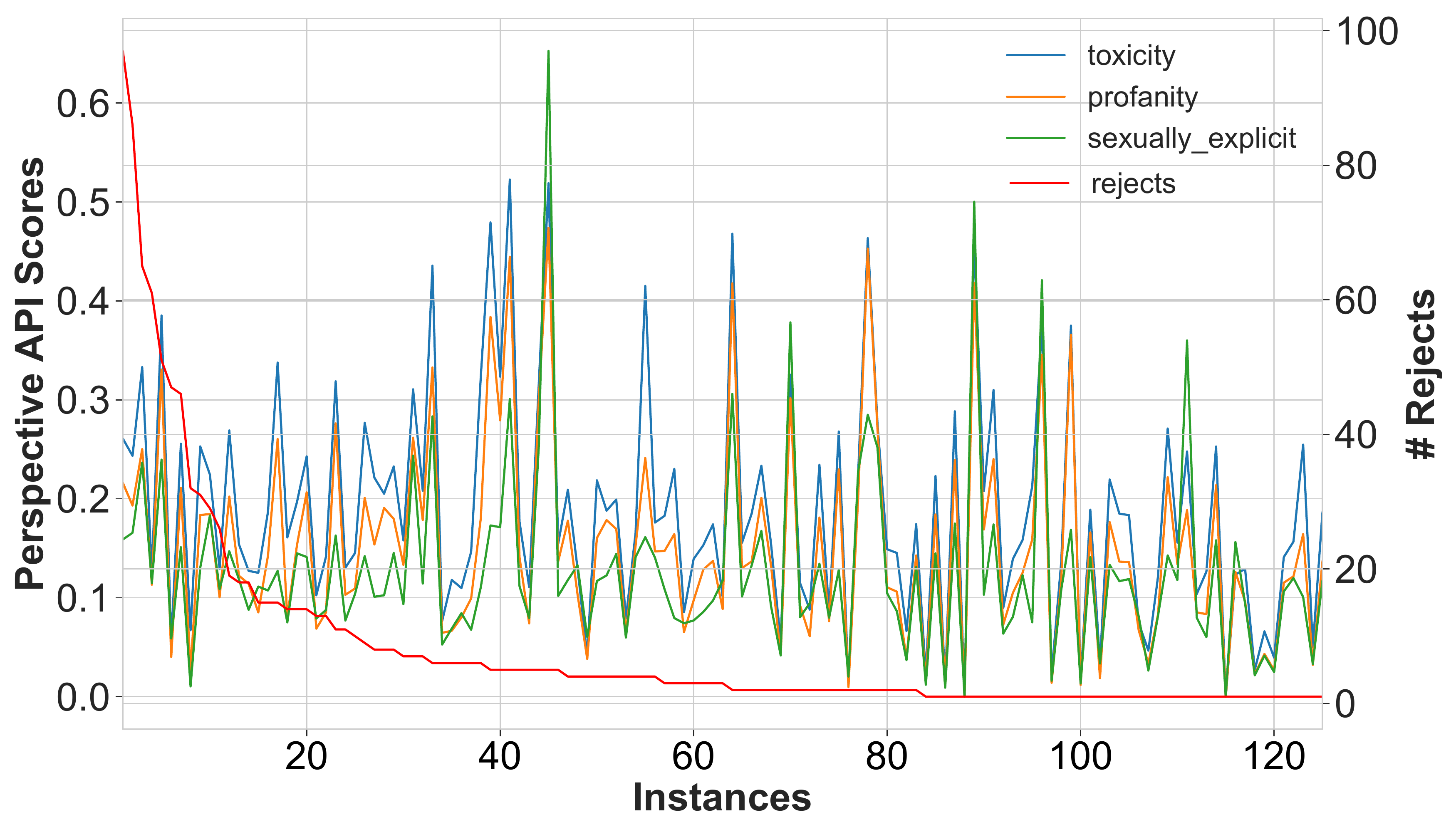}
  \caption{A plot of rejected Pleroma instances with the number of times they are rejected, their average toxicity, profanity, and sexually explicit scores across all users on the instances (sorted by the number of rejects).} 
  \label{fig:norm_rej}
\end{figure}

For context, Table \ref{table:rejected} lists the 5 most rejected Pleroma instances along with their number of users, posts, as well as their average scores in toxicity, profanity, and sexually explicit content.
Similarly, Figure~\ref{fig:norm_rej} shows a normalised plot with the Google Perspective API attribute features in Table \ref{table:rejected} for all the rejected Pleroma instances in our dataset. The score of each Perspective API attribute is the average score of all posts by users on an instance.
Although the instance with the most \texttt{reject} actions against it is \texttt{gab.com} (a Mastodon instance), Pleroma instances make-up 3 of the top 5.
Amongst the top 10 overall, just 40\% are from the Pleroma platform. This suggests a larger percentage of illicit material is imported into Pleroma from larger platforms such as \texttt{Mastodon} (probably due to the size of their user base).

\begin{table*}[]
\begin{tabular}{lrrrrrrl}
 \toprule
Instance & \begin{tabular}[c]{@{}r@{}}\#rejects\end{tabular} & \#users & \#user posts & Toxicity Sc & Profanity Sc & Sexually Explicit Sc  \\
\midrule
\begin{tabular}[l]{@{}l@{}}freespeech-extremist.com\end{tabular}     & 97                                                                  & 1.8k     & 1.13M      & 0.26 & 0.22 & 0.16 \\
kiwifarms.cc                 & 86                                                                  & 6.8k     & 391k        & 0.24 & 0.19 & 0.16 \\
spinster.xyz                 & 65                                                                  & 17.9k   & 1.34M       & NA & NA & NA \\
neckbeard.xyz                & 61                                                                  & 15.1k    & 816k        & 0.13 & 0.11 & 0.11 \\
poa.st                & 51                                                                  & 5.1k    & 344k        & 0.27 & 0.25 & 0.18 \\
\bottomrule
\end{tabular}

\caption{Top 5 Pleroma rejected instances with the number of times they are rejected, users, posts, the averages of their toxicity, profanity, and sexually explicit scores.}
 \label{table:rejected}
\end{table*}

\pb{Do rejected Pleroma instances retaliate?}
We next assess whether rejected Pleroma instances also tend to use the \texttt{reject} action themselves.
To answer this, we compute Spearman's correlation coefficient for rejects applied by \vs rejects received for all rejected Pleroma instances (-0.033). 
This means that the reverse is actually the case. In fact, we notice that the most rejected Pleroma instances barely apply the \texttt{reject} action against other instances (pleroma or non-Pleroma). For example, the most rejected Pleroma instance, \texttt{freespeechextremist.com}, does not reject a single other instance (Pleroma or non-Pleroma).
We conjecture that their openness to any kind of material may contribute to them being rejected.
Of the top 10, only 1 Pleroma instance (\texttt{spinster.xyz}, a woman-centric instance) has rejected over 2 instances (Pleroma and non-Pleroma) with 45 rejects. A manual check reveals non-tolerance for pornography, hate, violence or harassment in its ``Terms of Service''.

\pb{Why are instances blocked?.} 
To examine this, we manually annotate the 92 rejected Pleroma instances by going through their post content and also visiting each site. Note, these are the rejected Pleroma instances we have post data for, and we exclude single-user instances (see Section~\ref{section:coll}).
We label the rejected Pleroma instances as \one~Toxic (hate speech): for content with identity attacks, threats, insults, and other hateful material; 
\two~Sexually explicit (pornography): for adult content; 
\three~Profane: for material with swear/curse words; 
and \four~General: for Pleroma instances we are unable to categorise.
We are able to annotate 88.4\% of the rejected Pleroma instances.
For the Pleroma instances we are able to annotate, we find that the sexually explicit, toxic, and profane instances make up 90.6\%. The remaining 9.4\% are labeled as general.

\section{Is there collateral damage?}
\label{section:coll}

We conjecture that the activities of individual users may result in an entire instance being rejected.
Thus, many ``innocent'' users may also be rejected by association. We refer to this as \textbf{collateral damage}. To shed preliminary light on this question, we explore what fraction of users on rejected Pleroma instances share harmful material. 
We flag that there may be multiple reasons why an instance is rejected, and emphasize this limitation in our analysis.

\begin{figure}[t]
  \includegraphics[width=0.4\textwidth]{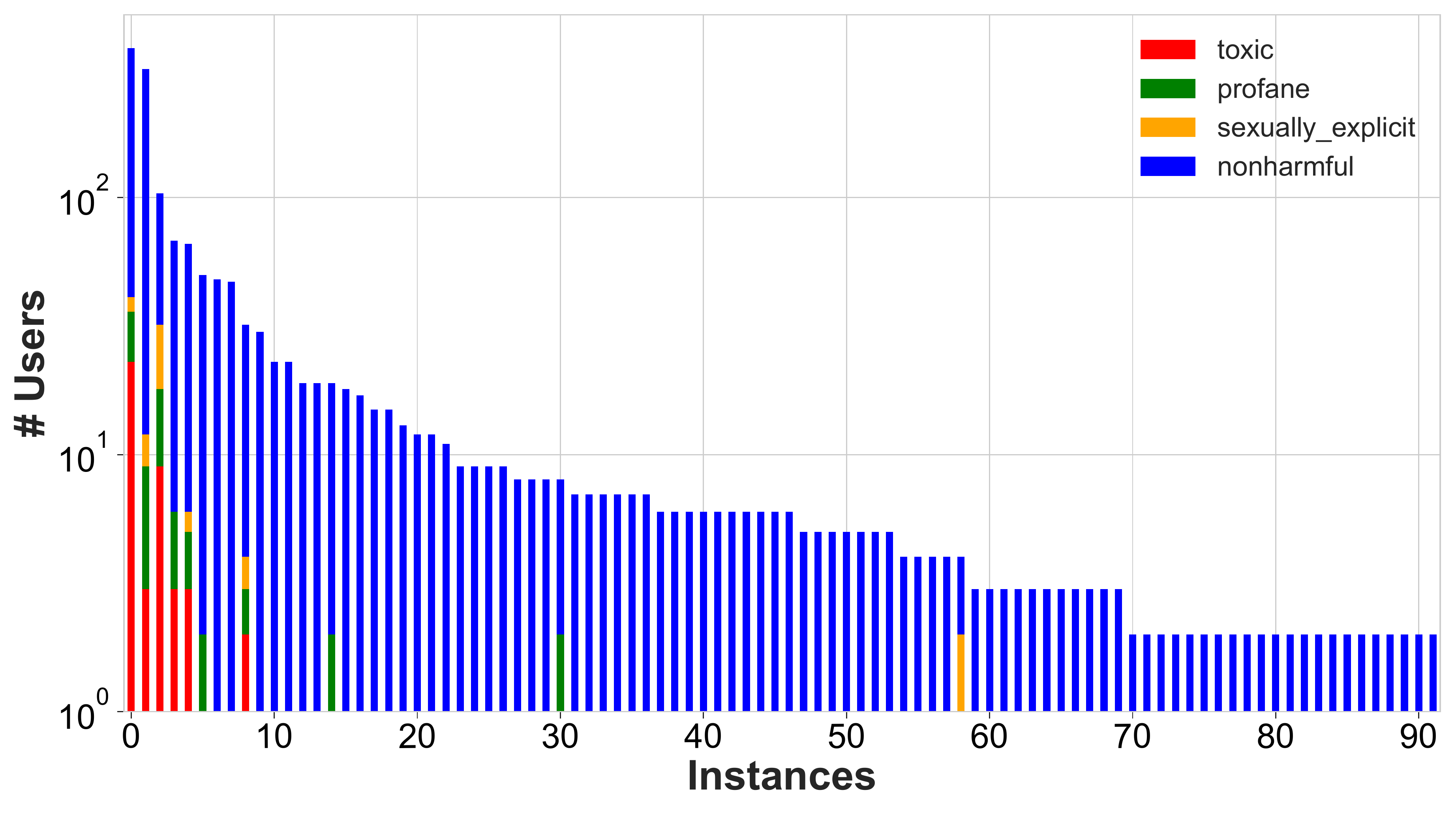}
  \caption{Number of toxic, profane, sexually explicit, and non-harmful users on each rejected Pleroma instance.}
  \label{fig:hate_nonhate}
\end{figure}

In total, we have posts for 61.9\% rejected Pleroma instances, with 26.4\% of them being single-user instances. 
As we are interested in what percentage of ``innocent'' users are affected by these policies, we filter out the single user Pleroma instances. 
We find 1.62k users on these rejected Pleroma instances have publicly accessible content (59.3k posts).
Using the Perspective labels (see Section~\ref{section:data}), we find that 4.2\% of these users on rejected Pleroma instances have a score of $>=$0.8 in at least one of the three attributes~\cite{corel}.
With our threshold of 0.8, we notice a harmful-to-non-harmful posts ratio of 1:11.
We also find that the Pleroma instances with posts averaging a Perspective API score >=0.8 
actually make up 7 of the top 10 most rejected Pleroma instances.

Figure~\ref{fig:hate_nonhate} shows a stacked bar plot, with the number of users that have been classified as toxic, profane, or sexually explicit on each rejected Pleroma instance. 
We also plot the number of non-harmful users on each Pleroma instance.
For the users with an average score >=0.8, we see a distribution of 69.7\% toxicity, 57.6\% profanity, and 43.9\% as sexually explicit.
 Note here that a user can be classified as all 3.
Based on this method, 95.8\% of the users on rejected Pleroma instances are affected, even though none of their own posts are flagged as harmful. This further strengthens our earlier hypothesis: it is likely that just a few posts from a few users trigger the rejections. 

For completeness, we finally show the percentage of harmful and non-harmful users when we use other Perspective thresholds in Table \ref{table:threshold}. We find that regardless of the threshold, a high percentage of non-harmful users are rejected alongside the small percentage of harmful users.

\begin{table}[t]
\setlength{\tabcolsep}{3pt}
\begin{tabular}{lcrrrr}
\toprule 
{\bf Threshold} & 0.5 & 0.6 & 0.7 & 0.8 & 0.9  \\
{\bf Non Harmful (\%)} & 86.4 & 91.8 & 94.1 & 95.8 & 97.3  \\

\bottomrule
\end{tabular}
\caption{Percentages of harmful and non-harmful Pleroma users with varying Google Perspective API thresholds.}
\label{table:threshold}
\end{table}

\section{Implications}

\pb{Collateral Damage.}
DW instance-based moderation provides useful tools for admins. However, our findings indicate that they are not granular enough.
\texttt{Reject} actions are applied to entire instances and all associated users are blocked. 
Despite this, we find that only 4.2\% of the users on rejected instances share harmful posts.
Although such users may be undesirable for other reasons, this does suggest that a large majority of users may be ``collateral damage''. 
This raises questions as to whether rejecting entire instances is the most appropriate approach. 

\pb{Dissatisfied Users.}
Generally, users tend to join social platforms where their friends already are, \ie the network effect~\cite{Efficient, behaviour}. This could possibly be one of the major reasons why some decentralised networks struggle to accumulate a strong user base. Being rejected could result in growing user dissatisfaction, especially if they have friends on another instance which they are banned from following. We argue that such experiences could lead to an exodus of users. Thus, addressing collateral damage is important. 

\pb{Federation Graph.}
In a centralised network, a social graph shows the connections between users. In the DW, this can go further to show the connections between instances. 
A \texttt{reject} tends to have far-reaching effects on the social graph; e.g., if an instance relies on another to reach a segment of the social graph, and due to the actions of a few users it gets rejected, that instance could be cut off from the wider network. This will adversely affect the federation. Exploring the wider impact is an interesting item for future work.

\section{Solution Space}

We now briefly explore some solutions to the above challenges. 
One obvious solution would be for administrators to adopt other less stringent in-built Pleroma policies, \eg tagging posts as NSFW. With this policy, messages from targeted instances are tagged with warnings rather than blocked. 

Some of the most rejected Pleroma instances happen to be those with sexually explicit content and these materials are mostly in media form. With the \texttt{media removal} facility, multimedia content is removed, leaving only the text.  That way, the harmful material loses its meaning while the non-harmful users are still able to have their posts delivered across the federation.
For DW platforms that carry out moderation in a similar fashion to Pleroma (\eg Mastodon), we concretely propose three steps that could be taken to improve moderation in the DW:
\begin{enumerate}[leftmargin=*]
    \item New generic policies could be designed that rely on a trusted/curated list of well-known instances in the fediverse that may need to be blocked. For example, policies called "\texttt{NoHate}" or "\texttt{NoPorn}" could have instances like \texttt{Gab, freespeechextremist.com} and \texttt{social.myfreecams.com}, and \texttt{baraag.net} listed as part of a community effort. Thus, an administrator could simply select the relevant lists. We expect that these listings are periodically updated by professionals who ensure that the instances have limited collateral damage.
    
    \item New user-driven policies could be designed that enable administrators to moderate on a per-user basis. For instance, \texttt{TagPolicy} can be used to apply a policy to a user based on a tag applied. Hence, streamline moderation interfaces could be devised to make the process of tagging individual users straightforward (potentially assisted by automated classifiers).
    
    \item New policies could enable administrators to automatically implement policies/actions for repeated offenders. For instance, policies/actions could be automatically applied (\eg NSFW, media removal) to a user when they have been reported $n$ times, or when the user post goes above a certain threshold (\eg in Google Perspective API). Again, such actions could be assisted by automatic classification of behaviour. 

\end{enumerate}

\section{Related Work}
\label{sec:related}

There has been a range of work on content moderation in social media.
Halevy et al.~\cite{integrity} looked into striking a balance between free speech and safety, considering diverse cultural and political climates.
Fortuna et al.~\cite{toxic} used 6 publicly available datasets and compared the labeling of each dataset for attributes such as sexism, toxicity, and racism. They found that definitions, datasets, and conflicting annotations can all affect the performance of classifiers.
Ribeiro et al.~\cite{hateful_users} studied differences between hateful Twitter users and normal users with respect to their activities, vocabulary and network centrality.
More recently, soft moderation and other techniques have also been explored by social platforms and researchers alike~\cite{zannettou2021won}.
Other studies have profiled social media users based on their dissemination of hate material~\cite{hate_beget_hate,spread_hate}. 
By contrast, our work focuses on the implementation of policies by administrators on Pleroma, rather than the behaviour of hateful users. 

There have also been a set of studies looking specifically at DW services. 
Raman et al. \cite{Mastodon} measured the challenges in deploying DW applications, particularly related to network issues~\cite{kaune2009modelling}.
Zignani et al. \cite{Matteo} studied the growth of Mastodon while comparing its structure with Twitter. Similarly, La Cava et al. \cite{fedi} explored the evolution of Mastodon at an instance level, as well as the connectivity between instances. Zignani et al. \cite{footprint} further investigated the interrelationship between the Mastodon system design and the social network.
Another closely related work \cite{content_warning} looked at how Mastodon users tag their own posts as NSFW.
Doan et al. \cite{dtube} investigated the performance of a decentralized video streaming platform (DTube) by developing an app that streams from both centralized and decentralized services.
We differ from these works in that we focus on content moderation activities.

\vspace*{-0.2cm}
\section{Conclusion}

This paper presented the first study of Pleroma and its moderation policies. We find that policies are widely used, impacting 97.7\% of users and 97.8\% of posts. 
Using the Perspective API, we have found that 95.8\% of the users on rejected Pleroma instances do not share posts classified as harmful.
This leaves just 4.2\% of harmful users that are likely responsible for the rejects. This implies significant ``collateral damage.'' This has led us to sketch a set of strawman policies that may reduce this damage. Our proposed solutions would generally be applicable to platforms that carry out moderation at a per-instance granularity (\eg DW) rather than at a per-user granularity (\eg  Twitter). In our future work, we plan to implement and evaluate these policies, as well as continue to explore how other DW platforms perform moderation. 

There are a number of lines of future work. We wish to further explore the reasons why administrators apply particular policies. This, for instance, could be achieved via user studies. 
Using this knowledge, we intend to develop novel policies that can assist administrators. We are particularly interested in designing more techniques that can automatically identify users or instances to apply certain policies too. 

\noindent\textbf{Acknowledgements.}
This work is supported by EU H2020 grant agreements No 871793 (Accordion), No 871370 (Pimcity), and No 101016509 (Charity), as well as EPSRC EP/S033564/1 grant and the UK's National Research Centre on Privacy, Harm Reduction, and Adversarial Influence Online (REPHRAIN, UKRI grant: EP/V011189/1).

\bibliographystyle{abbrv}

\appendix

\section{Appendix}
We summarise the basic functionalities of Pleorma in-built policies in Table \ref{table:pol}. 
\label{append:pol}
\begin{table*}
\resizebox{2\columnwidth}{!}{
\begin{tabular}{llrl}
 \toprule
policy & \begin{tabular}[c]{@{}r@{}}\#description\end{tabular} & \#instances & \#users  \\
\midrule
 ObjectAge &  Rejects or delists posts based on their age when received & 869 & 57,854\\
                                 TagPolicy & Applies policies to individual users based on tags &  429 & 38,067\\
                              SimplePolicy & Restrict the visibility of activities from certains instances with a suite of actions &  330 & 46,691\\
 NoOpPolicy & Doesn’t modify activities (default) &  176 & 6,443\\
                        HellthreadPolicy & De-list or reject messages when the set number of mentioned users threshold is exceeded &  87 & 14,401\\
                        StealEmojiPolicy &  List of hosts to steal emojis from & 81 & 7,003\\
                        HashtagPolicy &  List of hashtags to mark activities as sensitive (default: nsfw) & 62 & 10,933\\
                        AntiFollowbotPolicy &  Stop the automatic following of newly discovered users & 51 & 6,918\\
                        MediaProxyWarmingPolicy &  Crawls attachments using their MediaProxy URLs so that the MediaProxy cache is primed & 46 & 9,851\\
                        KeywordPolicy &  A list of patterns which result in message being reject/unlisted/replaced & 42 & 22,428\\
                        AntiLinkSpamPolicy & Rejects posts from likely spambots by rejecting posts from new users that contain links &  32 & 7,347\\
                        ForceBotUnlistedPolicy & Makes all bot posts to disappear from public timelines &  23 & 6,746\\
                        EnsureRePrepended & Rewrites posts to ensure that replies to posts with subjects do not have an identical subject and instead begin with re: &  18 & 247\\
                        ActivityExpirationPolicy & Sets a default expiration on all posts made by users of the local instance &  11 & 1,420\\
                        SubchainPolicy & Selectively runs other MRF policies when messages match  &  8 & 81\\
                        MentionPolicy & Drops posts mentioning configurable users  &  6 & 1,149\\
                        VocabularyPolicy & Restricts activities to a configured set of vocabulary  &  5 & 121\\
                        AntiHellthreadPolicy & Stops the use of the HellthreadPolicy  &  4 & 2,106\\
                        RejectNonPublic & Whether to allow followers-only/direct posts  &  3 & 1,101\\
                        FollowBotPolicy & Automatically follows newly discovered users from the specified bot account  &  2 & 281\\
                        DropPolicy                      & Drops all activities &  1 & 1,098\\

\bottomrule
\end{tabular}
}

\caption{Description of policies provided by Pleroma and the number of instances that enable them, as well as the number of users on the instances}

 \label{table:pol}
\end{table*}

\begin{figure*}[!h]
	\includegraphics[width=\linewidth]{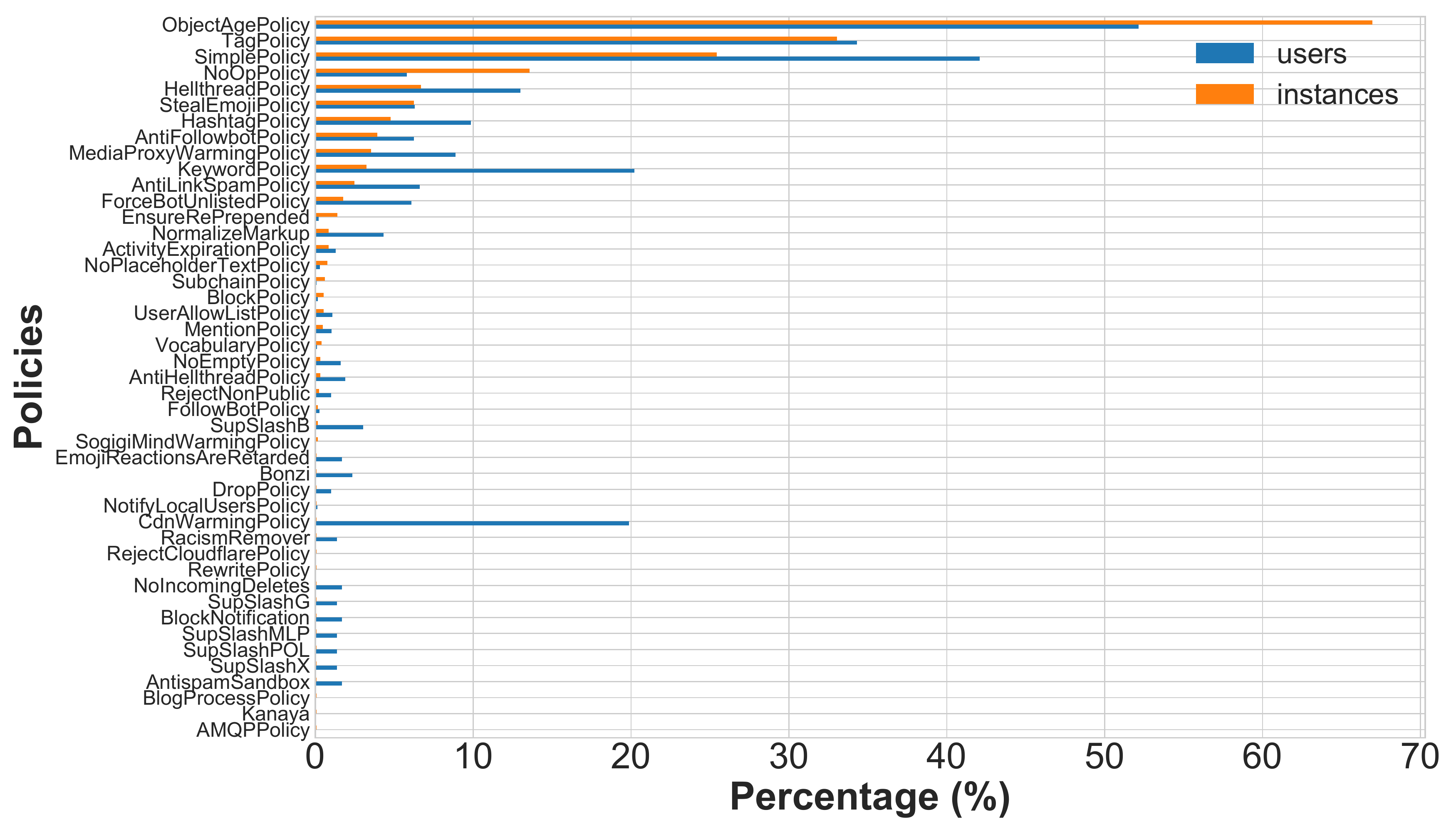}
	\caption{Entire policy spectrum showing policy types and the percentage of instances that use each policy (sorted by the percentage of instances). We also include the percentage of the global user population on the instances that use each policy.}
	\label{fig:all_pols}
\end{figure*}

For completeness, we show in Figure~\ref{fig:all_pols} the entire policy spectrum, the percentage of Pleroma instances that enable these policies, as well as the number of users on these instance.

\end{document}